\documentclass[preprint,showpacs,preprintnumbers,amsmath,amssymb]{revtex4} % newly added
\usepackage{amssymb}
\usepackage{graphicx}% Include figure files
\usepackage{dcolumn}% Align table columns on decimal point
\usepackage{bm}% bold math
\usepackage{booktabs}

\begin{document}

\title{Efficient Quantum Ratchet}
\author{Chuan-Feng Li$\footnote{email: cfli@ustc.edu.cn}$, Rong-Chun Ge, Guang-Can Guo}
\affiliation {Key Laboratory of Quantum Information, University of
Science and Technology of China, CAS, Hefei, 230026, People's
Republic of China}
\date{\today }
%\pacs{68.65.Hb, 73.22.-f, 78.67.Hc }% PACS, the Physics and Astronomy
% 68.65.-k Low-dimensional, mesoscopic, and nanoscale systems:
%          structure and nonelectronic properties
% 68.65.Hb Quantum dots
% 73.22.-f Electronic structure of nanoscale materials: clusters,
%          nanoparticles, nanotubes, and nanocrystals
% 73.21.La Electron states and collective excitations in multilayers, quantum
%          wells, mesoscopic, and nanoscale system: Quantum dots
% 78.67.Hc Optical properties of low-dimensional structures: Quantum dots
% 73.22.-f Electronic structure of nanoscale materials: clusters,
%          nanoparticlehell, nanotubes, and nanocrystals
% 71.15.-m computational methodology use 71.15.-m
%nanoparticles, nanotubes, and nanocrystals

\begin{abstract}
Quantum resonance is one of the main characteristics of the quantum
kicked rotor, which has been used to induce accelerated ratchet
current of the particles with a generalized asymmetry potential.
Here we show that by desynchronizing the kicked potentials of the
flashing ratchet [Phys. Rev. Lett. 94, 110603 (2005)], new quantum
resonances are stimulated to conduct directed  currents more
efficiently. Most distinctly, the missed resonances $\kappa=1.0\pi$
and $\kappa=3.0\pi$ are created out to induce even larger currents.
At the same time, with the help of semiclassical analysis, we prove
that our result is exact rather than phenomenon induced by errors of
the numerical simulation. Our discovery may be used to realize
directed transport efficiently, and may also lead to a deeper
understanding of symmetry breaking for the dynamical evolution.

\end{abstract}


\maketitle

Extraction of work from a system without macroscopic bias is a topic
of interest all the time, and it has drawn attentions of researchers
from all kinds of fields \cite{Rei02,Ha09,Ro11}. Ratchet effect, the
phenomenon of creating out directed current from periodic
configuration without macroscopic bias under the situation of
symmetry breaking, offers an effective way of extracting work. In
order to display directed current, the system must be driven out of
equilibrium, and obtain a broken symmetry. Depend on different kinds
of working mechanism, there are dissipative ratchets and Hamiltonian
ratchets. For the dissipative ratchets, noise play an indispensable
role \cite{As02,Ha05}. While for the latter, where noise is absent,
the ratchet currents may be induced by chaos \cite{Ju96}. The
quantum version of ratchet effect has also attracted intensive
attention in the past few decades. Most of the investigations of
quantum ratchet effect concentrate on open quantum system suffering
from noise \cite{Lin99}, and current reversal due to quantum effect
has been shown.

Due to the advance in optical lattice, coherent quantum ratchets
realized with cold atoms \cite{Jon04,Jon07} and Bose-Einstein
condensate \cite{Da08,Sal09} have been a new spot of research
\cite{Mon02,Gon06,Gon07,Chen09}. In addition, the effect of
nonlinear interaction among the Bose-Einstein condensate on its
dynamics has been concerned \cite{Zh04,Mon09}. It was first shown in
Ref. \cite{Lun05} that low order quantum resonances could be used to
create out accelerated ratchet currents in the quantum flashing
ratchet, and later high order quantum resonances were found to
induce even larger ratchet currents as the effective strength of the
flashing potential increase \cite{Ken08}. The quantum flashing
ratchet is a generalization of the quantum kicked-rotor. As a
paradigm for quantum chaos, the quantum kicked-rotor with symmetry
potential has been explored intensively since 1980s
\cite{Ca79,Iz80,Sh82,Fi82,Ch86,Mo95,Le08}. It was found that the
quantum kicked-rotor will display quantum resonance when the ratio
between the effective Plank number $\kappa$ and $4\pi$ is a rational
number, at which the average energy of the system over time is
proportional to the square of the kicked numbers, and dynamical
localization when $\kappa$ is a general value, at which the average
energy over time becomes saturated after a few periods.

In this paper we study the transport of a particle driven by delta
kicks in one dimensional optical lattice much like the flashing
ratchet demonstrated in Ref. \cite{Jon04}. We show that by
desynchronizing the two standing laser waves employed in Refs.
\cite{Lun05,Ken08}, quantum resonances can be stimulated to give
rise to directed currents more efficiently. By numerical simulation,
we find that when the time delay between the two symmetry potentials
($\sin(2x)$ and $\sin x$) equals half of the period, new quantum
resonances are stimulated to induce ratchet currents. The low order
quantum resonances $\kappa=\pi$ and $3\pi$ missed out in Refs.
\cite{Lun05,Ken08} have been created out to induce even larger
currents. We also make sure of our numerical results with the help
of semiclassical analysis at quantum resonance $\kappa=1.0\pi$. What
is more, high order quantum quantum resonances can be stimulated
with even weaker strength of the potential, which may be a piece of
good news for directed transport in biological system.

In dimensionless units, the model we concern about is a
generalization of the flashing ratchet \cite{Lun05}, which is
described by
\begin{eqnarray}
{\rm i}\kappa\frac{\partial\psi}{\partial
t}&=&-\frac{\kappa^2}{2}\frac{\partial^2\psi}{\partial
x^2}+K\big[v_1(x)\sum_{n=0}^{\infty}\delta(t-n-\eta)\nonumber\\
&+&v_2(x)\sum_{n=1}^{\infty}\delta(t-n)\big],
\end{eqnarray}
with $v_1(x)=\alpha\sin(2x)$ and $v_2(x)=\sin x$. $\alpha$ denotes
the relative strength of the two periodic potentials realized with
two standing laser waves, whose spatial periodicity are $\lambda/4$
and $\lambda/2$ ($\lambda=2\pi/k_L$) respectively. $0\leq1-\eta<1$
denotes the time delay between the two periodic potentials.
$\kappa=8\omega_R T$ is the effective Plank constant, which defines
the quantum nature of the system (quantum resonance and dynamical
localization), where $\omega_R=\kappa k^2_L/2m$ is the recoil
frequency of the applied field with periodicity $\lambda$, and $T$
is the period of the flashing potential with the same frequency.
$K=\frac{\kappa T V_0}{\hbar}$ is the effective strength of the
potential, where $V_0$ is the amplitude of the potential induced by
the kicked filed. To make the following expression more convenient,
we also define $P=K/\kappa$.

The time evolution of the system in one period is composed of a
succession of free evolution with the total time interval 1
separated by two sequences of flashing kicked potentials, and is
given by
\begin{eqnarray}
|\psi(t+1)\rangle&=&e^{-{\rm i}Pv_2(x)}e^{-{\rm i}(1-\eta)\kappa\hat{k}^2/2}\nonumber\\
&\times&e^{-{\rm i}Pv_1(x)}e^{-{\rm
i}\eta\kappa\hat{k}^2/2}|\psi(t)\rangle,
\end{eqnarray}
where $\hat{k}=-{\rm i}\partial/\partial x$ is the effective
momentum operator, and the wave function shown here is valid only at
integer time $t$.

As we have noted above, in order to get directed currents, symmetry
breaking is needing. Here, in order to remove the effect of symmetry
breaking in the dynamical evolution induced by the initial
condition, we concentrate on the uniform zero momentum state
$\psi(x, t=0)=\frac{1}{\sqrt{2\pi}}$, which is also a good
approximation of the ultra-cold atoms loaded on optical lattice
spreading over several periods. At the same time, we take
$\alpha=0.3$ throughout the whole paper.

\begin{figure}
\includegraphics[width=3in]{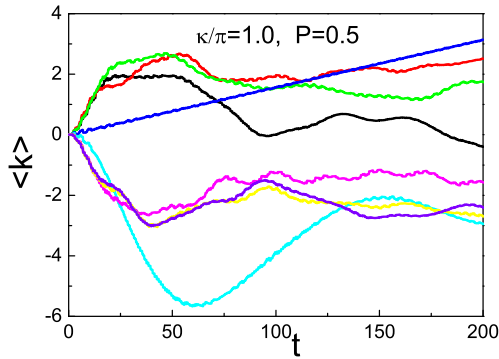}
\caption{(Color online) The dependence of ratchet current over time
for different time delays $\eta$ at quantum resonance
$\kappa=1.0\pi$, with $\eta=1/7,2/7,3/8,1/2,5/9,2/3,7/10,4/5$
represented by the black, red, green, blue, cyan, magenta, and
violet lines respectively.}
\end{figure}

As has been shown in Refs. \cite{Lun05,Ken08}, we find that there
are no accelerated ratchet currents for general values of $\eta$ at
the quantum resonance $\kappa=1.0\pi$. But there are ratchet
currents at short time interval. What is more, the directions of the
currents depend on the time delay between the two sequences of
symmetry kicked potentials. For little value of $\eta$, it tends to
develop current parallel with the $x$ axis in the first few tens of
kicks, while for large value of $\eta$, it tends to get reverse
current, as in shown in Figure 1, with the effective strength of the
potential $P=0.5$. It indicates that, with different time delays,
there are different degrees of imbalance between the left and right
sides.

The result is unique at the time delay $\eta=1/2$, as is shown in
Fig. 1 that accelerated ratchet current emerges. At the macro-level,
the accelerated ratchet current signifies considerable asymmetry
between different directions. Here, the broken symmetry is the
result of desynchronizing of the two time sequences of symmetry
potentials, and the degree of asynchronization can be accessed by
the quantity
$\textsf{min}\{\frac{\eta}{1-\eta},\frac{1-\eta}{\eta}\}$. Then, at
least at the quantum resonance $\kappa=1.0\pi$, time
asynchronization can lead to the symmetry breaking of dynamical
evolution more efficiently than that induced by instantaneously
potentials with broken symmetry in the position space. As can be
seen in Fig. 2(a), due to the time delay $\eta=0.5$, the ratchet
current created out at $\kappa=1.0\pi$ exceeds that at
$\kappa=0.5\pi$ even with a weak potential strength $P=0.5$. In
addition, as the time delay $\eta$ increase from 0 to 0.5, the
ratchet current at $\kappa=0.5\pi$ becomes larger too. But as
demonstrated in Fig. 2(b), the atom with $\eta=0.5$, cannot absorb
energy as efficiently as it with $\eta=0$. Fig. 2(c) shows the
currents with different strength of potentials over 200 periods. The
ratchet current stimulated out of quantum resonance $\kappa=1.0\pi$
with time delay $\eta=0.5$ performs much better than that out of
$\kappa=0.5$ without a too much strong potential, so it can be used
as new operating point to realize directed transport.

\begin{figure}
\begin{center}
\includegraphics[width=3in]{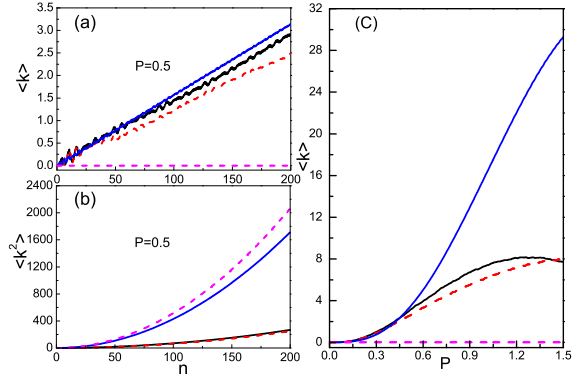}
\end{center}
\caption{(Color online) The development of (a) the ratchet currents
$<k>$ and (b) effective energies $<k^2>$, with the strength of
potential equals $P=0.5$. (c) Ratchet currents $<k>$ with different
values of the potential strength over 200 periods. Here, the dark
solid line represents $\kappa=0.5\pi,\eta=0.5$, red dashed line
$\kappa=0.5\pi,\eta=0.0$, blue solid line $\kappa=1.0\pi,\eta=0.5$,
and magenta dashed line represents $\kappa=1.0\pi,\eta=0.0$.}
\end{figure}

It is well known that all kinds of properties of the Hamiltonian
system are fixed by its energy spectrum. From the microscopic
observation, Floquet analysis \cite{Lun06} can be exploited to deal
with our system due to the fact that the system is subjected to time
periodic interactions. The quasienergy spectra of our system at
quantum resonance $\kappa=1.0\pi$ and time delay $\eta=0.5$ are
given by
\begin{eqnarray}
\omega^{1,3}_{x_0}&=&\pi/4\mp S(x_0)\nonumber\\
\omega^{1,3}_{x_0}&=&\pi/4\mp \overline{S}(x_0)
\end{eqnarray}
here $S(x)$ and $\overline{S}(x)$ are given by
\begin{eqnarray}
S(x_0)&=&\arctan\bigg[\frac{\sqrt{1+\frac12\cos^2(\overline{v_2}(x_0))q_1(x_0)}}{\cos(\overline{v_2}(x_0))p_1(x_0)}\bigg]\nonumber\\
\overline{S}(x_0)&=&\arctan\bigg[\frac{\sqrt{1+\frac12\cos^2(\overline{v_2}(x_0+\frac{\pi}{2}))q_2(x_0)}}{\cos(\overline{v_2}(x_0+\frac{\pi}{2}))p_2(x_0)}\bigg]
\end{eqnarray}
with $p_1(x_0)=\cos\big(\overline{v_1}(x_0)+\frac{\pi}{4}\big)$,
$p_2(x_0)=\cos\big(\overline{v_1}(x_0)-\frac{\pi}{4}\big)$,
$q_1(x_0)=\sin\big(2\overline{v_1}(x_0)\big)-1$, and
$q_2(x_0)=\sin\big(2\overline{v_1}(x_0)\big)+1$. In order to make
our expression more compact we have used $\overline{v_i}(x)$ in
place of $Pv_i(x)$, with $i=1,2$.

As is shown in Fig. 3(b), due to the time delay, the quasienergy
spectra are split into four bands, with bands $\omega^1_{x_0}$
($\omega^2_{x_0}$) and $\omega^3_{x_0}$ ($\omega^4_{x_0}$) intersect
at some position of $x_0$. While for the case of $\eta=0$
\cite{Lun05,Ken08,Lun06}, as is shown in Fig. 3(a), things are quite
different, where there are only two bands of quasienergy spectra
without crossing. The corresponding eigenvectors $\alpha^{\rm i}$ at
$\eta=0.5$ with ($i=1,2,3,4$) are given by
\begin{eqnarray}
\alpha^{1,3}_{x_0,0}&=&\frac{p_2(x_0)e^{-{\rm
i}\big(\overline{v_2}(x_0)\pm
S(x_0)\big)}}{\sqrt{2-2p_1(x_0)\cos\big(\overline{v_2}(x_0)\pm
S(x_0)\big)}},\nonumber\\
\alpha^{1,3}_{x_0,2}&=&\frac{-{\rm i}\big[1-p_1(x_0)e^{-{\rm
i}\big(\overline{v_2}(x_0)\pm
S(x_0)\big)}\big]}{\sqrt{2-2p_1(x_0)\cos\big(\overline{v_2}(x_0)\pm
S(x_0)\big)}},\nonumber\\
\alpha^{1,3}_{x_0,1}&=&\alpha^{1,3}_{x_0,3}=0.
\end{eqnarray}
and
\begin{eqnarray}
\alpha^{2,4}_{x_0,1}&=&\frac{p_1(x_0)e^{-{\rm
i}\big(\overline{v_2}(x_0+\frac{\pi}{2})\pm
\overline{S}(x_0)\big)}}{\sqrt{2-2p_2(x_0)\cos\big(\overline{v_2}(x_0+\frac{\pi}{2})\pm
\overline{S}(x_0)\big)}},\nonumber\\
\alpha^{2,4}_{x_0,3}&=&\frac{-{\rm i}\big[1-p_2(x_0)e^{-{\rm
i}\big(\overline{v_2}(x_0+\frac{\pi}{2})\pm
\overline{S}(x_0)\big)}\big]}{\sqrt{2-2p_2(x_0)\cos\big(\overline{v_2}(x_0+\frac{\pi}{2})\pm
\overline{S}(x_0)\big)}},\nonumber\\
\alpha^{2,4}_{x_0,0}&=&\alpha^{2,4}_{x_0,2}=0,
\end{eqnarray}

\begin{figure}
\begin{center}
\includegraphics[width=2.5in]{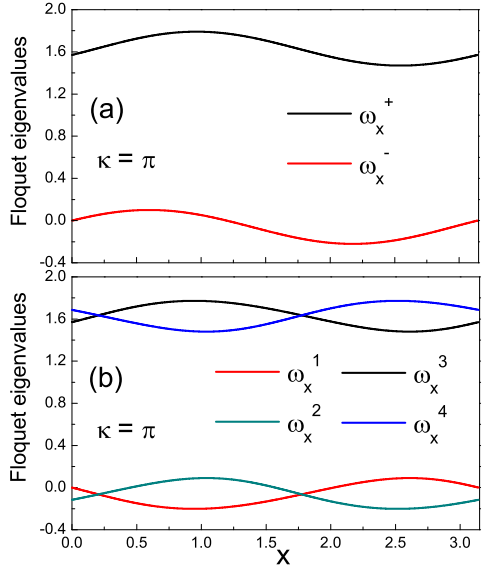}
\end{center}
\caption{(Color online) Floquet eigenvalues as functions of position
$x$, with the strength of potential equals $P=0.5$, at quantum
resonance $\kappa=1.0\pi$. (a) The flashing ratchet without time
delay $\eta=0$ \cite{Lun05}, has two bands of quasienergy spectra.
(b) The flashing ratchet with time delay $\eta=1/2$, has four bands
of quasienergy spectra.}
\end{figure}

With the help of these basics elements, the wave function of the
system at integer time reads
\begin{equation}
\psi(x,t)=\frac{1}{\sqrt{2\pi}}\sum_{\mu,l,k}\alpha^{\mu*}_{x-l\pi/2,k}e^{-{\rm
i}\omega^{\mu}_{x-l\pi/2}t}\alpha^{\mu}_{x-l\pi/2,l}.
\end{equation}
The wave function at half integer time can be obtained from the wave
function at integer time by reversing the action of the second
flashing potential $v_2(x)$ and the free evolution of the system
with time interval 0.5, so it is given by
\begin{eqnarray}
\psi(x,t-\frac12)&=&\langle x|e^{{rm i}\pi \hat{k}^2/4}e^{{\rm
i}\overline{v_2}(x)}|\psi(x,t)\rangle\nonumber\\
&=&\sum_{k=-\infty}^{\infty}e^{{\rm i}\pi k^2/4-{\rm
i}k(x-x')}e^{{\rm i}\overline{v_2}(x')}\psi(x,t).
\end{eqnarray}
It can be shown straightforward through a simply numerical
simulation that at the limit of large value of $t$, the average
force exerted on the atom over one period is a non zero constant. So
under the situation of semiclassical, we can now make sure that our
numerical results at quantum resonance $\kappa=1.0\pi$ and time
delay $\eta=0.5$ demonstrate the real accelerated directed currents.
The formulation of the force can be given by
\begin{eqnarray}
&&\langle\psi(t-\frac12)|\frac{d\overline{v_1}(x)}{dx}|\psi(t-\frac12)\rangle+\langle\psi(t)|\frac{d\overline{v_2}(x)}{dx}|\psi(t)\rangle\nonumber\\
&\propto&\int_0^{2\pi}dx\bigg(\psi^*(t-\frac12)\frac{d\overline{v_1}(x)}{dx}\psi(t-\frac12)\nonumber\\
&+&\psi^*(t)\frac{d\overline{v_2}(x)}{dx}\psi(t)\bigg).
\end{eqnarray}

In fact, we also find that when we reverse the order of the two
symmetry potentials $v_1(x)$ and $v_2(x)$, there is hardly any
difference. For $P=1.0$ the relative difference measured by
$|2\times\frac{<k_1>-<k_2>}{<k_1>+<k_2>}|$ over 200 periods is 0.7
percent, and for $P=3.0$, it is 0.4 percent (Here $<k_1>$ is the
average current when $v_1(x)$ is first exerted, and $<k_2>$ is the
average current when $v_2(x)$ is first exerted). Current reverse is
found around $P=2.6$ for both cases.

It is found that, at time delay $\eta=0.5$, high order quantum
resonances can be stimulated to lead  to accelerated ratchet
currents more efficiently than those at $\eta=0$. An example in
point is quantum resonance $\kappa=2.625\pi$, it can induce a
ratchet current even at $P=1.5$ much smaller than that without time
delay. Another unique low order quantum resonance is
$\kappa=3.0\pi$, it can develop directed current as efficiently as
$\kappa=1.0\pi$. So it could be valuable for directed transport in
biological system, in which the external potential imposed on the
organism should be never too strong.

\begin{figure}
\begin{center}
\includegraphics[width=4in]{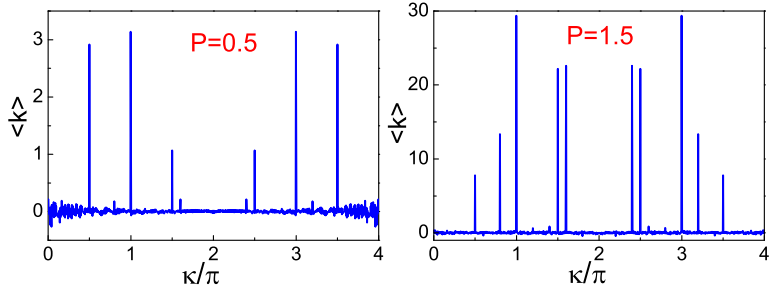}
\end{center}
\caption{(Color online) Ratchet currents measured by the mean wave
number $<k>$ as functions of effective Plank constant $\kappa$ after
200 periods, with $\eta=0.5$. (a) At strength of potential $P=0.5$.
(b) At strength of potential $P=1.5$.}
\end{figure}

In Fig. 4, we show the average wave number $<k>$ over 200 periods,
with time delay $\eta=0.5$ for different values of effective Plank
constant $\kappa$. As is shown, when the potential strength $P$
increase from 0.5 in Fig. 4(a), to 1.5 in Fig. 4(b), almost all the
quantum resonances lead to larger ratchet currents. Higher order
quantum resonances begin to surpass the low order ones as the
potential strength increase. But quantum resonances $\kappa=1.0\pi$
and $\kappa=3.0\pi$ still mark themselves with large ratchet
currents, which are the result of the time asymmetry among the
flashing kicks. The other resonances are correlated with both
configuration space asymmetry and time series asymmetry whose
effects may not be always in phase. We also find that when we
reverse the time order of $v_1(x)$ and $v_2(x)$, there is hardly any
difference.

In conclusion, in this paper we have studied the ratchet effect by
desynchronizing the two symmetry kicked potentials. It is found that
with a time delay $\eta=0.5$ between the two symmetry potentials,
quantum resonances $\kappa=1.0\pi$ and $\kappa=3.0\pi$ are
stimulated to lead much larger ratchet currents than other quantum
resonances at weak potential strength, so they can be chosen as new
operating points in order to exploit directed transport. The ratchet
effect of $1.0\pi$ ($3.0\pi$) is the result of time asymmetry of the
two kicked series of potentials. At this point, time asymmetry may
lead to symmetry breaking of the dynamical evolution between
different directions more efficiently than that with asymmetry in
the position space at discrete times. At the same time, we find that
high order quantum resonances can be created out to lead ratchet
currents more efficiently than those without a time delay.

This work was supported by National Fundamental Research Program,
National Natural Science Foundation of China (Grant Nos.60921091,
10874162).

\end{document}